\documentclass[11pt]{article}

\advance \textheight by \topskip
\usepackage{amsmath,amsfonts}
\usepackage[english]{babel}
 \usepackage{color}
\usepackage{amsfonts,dsfont}
\usepackage{amsmath,amssymb}
\usepackage{mathrsfs}
\usepackage{stmaryrd}
\usepackage[english]{babel} 
\usepackage[bookmarksnumbered,
plainpages=false,colorlinks=true,linkcolor=black,
citecolor=black,urlcolor=blue]{hyperref}

\begin{document}

\title{Higher-spin symmetries of the free Schr\"odinger equation}

\author{{\sf Mauricio Valenzuela}\footnote{mauricio.valenzuela@uss.cl} \\[8pt]
 {\small \it Facultad de Ingeniería y Tecnolog\'ia}\\ 
{\small \it Universidad San Sebasti\'an, General Lagos 1163, Valdivia 5110693, Chile
}}
 
\date{
}

\maketitle

\begin{abstract}
  It is shown that the Schr\"odinger symmetry algebra of a free particle in $d$
spatial dimensions can be embedded into a representation of the higher-spin
algebra. The latter spans an infinite dimensional algebra of
higher-order symmetry generators of the free Schr\"odinger equation. An explicit
representation of the maximal finite dimensional subalgebra of the higher spin
algebra is 
given in terms of non-relativistic generators. We show also how to convert 
Vasiliev's equations into an explicit non-relativistic covariant
form, such that they might apply to non-relativistic systems. Our procedure
reveals that the space of solutions of the Schr\"odinger equation can
be regarded also as a supersymmetric module.
 \end{abstract}

\bigskip

\begin{center}
\end{center}

\maketitle

\bigskip

\section{Introduction}

The Schr\"odinger group was discovered by S. Lie \cite{Lie} and, even earlier,
the conserved quantities associated with the Schr\"odinger invariance were
already known to Jacobi \cite{Jacobi} (see also \cite{Plebanski,Duval:2009vt}).
Its name, however, is taken from quantum mechanics
\cite{Kastrup,Jackiw,Nie,Nie2,Hag}, since it extends the Galilean symmetries of
the free Schr\"odinger equation with dilatation and expansion transformations.
The Schr\"odinger group is isomorphic to the Newton-Hooke conformal group of the
harmonic oscillator \cite{Nie2}, and it appears also in several contexts, e.g.,
magnetic monopoles \cite{Jackiw:1980mm}, vortices
\cite{Jackiw:1989qp,Jackiw:1990mb,Duval:1994pw,Horvathy:2008hd}, fluid mechanics \cite{fluid},
and strongly correlated fermions \cite{Henkel:1993sg,Mehen:1999nd}. The
Schr\"odinger group has some infinite dimensional generalizations
\cite{Plebanski,Henkel:1993sg}, and it can be also realized geometrically as
space-time isometries \cite{Henkel:2003,Duval:1990hj,Duval:2008jg}. The
Schr\"odinger symmetry have attracted also renewed interest in the context of non-relativistic $AdS/CFT$ correspondence 
\cite{Son:2008ye,Balasubramanian:2008dm,Maldacena:2008wh}, which relates an
asymptotic theory on a curved background to a non-relativistic quantum system
\cite{Duval:2008jg,Son:2008ye,Balasubramanian:2008dm,Maldacena:2008wh}. This is
a consequence of the embedding of the Schr\"odinger algebras into relativistic
conformal algebras \cite{Burdet}.

The goal of this letter is to show that Galileo boosts, translations and the
mass generators  are
building blocks for constructing higher-order symmetries of the free
Schr\"odinger equation. This is done by noticing that from Galileo boosts,
translations and the mass operator, which satisfy the Heisenberg algebra, we can
construct a representation of the Weyl algebra. The latter will span an infinite
set of conserved charges, containing in particular all Schr\"odinger
generators. This can be regarded as the non-relativistic analog
of the Eastwood result \cite{East} on the maximal symmetries of the massless
Klein-Gordon equation, which is spanned by polynomials in 
conformal symmetry generators.

Endowing the generators of the Weyl algebra with a (super)commutator
product yields the so-called \textit{higher spin (super)algebras}
\cite{Vasiliev1,Vasiliev2,Vasiliev3,Vasiliev3b}. These algebras are well-known in the
context of higher spin gauge theory (see also
\cite{Vasiliev4,Blencowe,Boulanger,Henneaux}). It follows that {the free
Schr\"odinger equation exhibits higher spin symmetries}, and reciprocally,
{Schr\"odinger symmetry is naturally contained in higher spin theory}. As
we shall
see at the end of this paper, Vasiliev equations of
higher spin gauge fields can be written in an explicit non-relativistic
form, by simple identification of the symplectic-spinor indices of the higher
spin fields with spatial indices of a non-relativistic space-time. Indeed, the
truncation of the higher spin algebra to its maximal finite dimensional
subalgebra contains the Schr\"odinger algebra. Reciprocally, the correspondence
between symplectic-spinor indices and non-relativistic spatial indices allows
 to endow the Sch\"odinger generators with a
supercommutator product, yielding an orthosymplectic-type supersymmetry of the
Schr\"odinger equation. 

\section{Symmetries of the Schr\"odinger equation}

Consider  $d$ dimensional Heisenberg algebra,  
\begin{equation}
\mathfrak{h}_d=\{\mathds{1},\,G_i,\; P_i\; \; ; \quad [G_i,P_j]={\rm i}\delta_{ij} \; , \; \; i , j \ = 1,...,d
 \}.\nonumber
\end{equation}

The Weyl algebra, denoted $\mathfrak{h}_d^*$, can be defined as the algebra of
Weyl ordered polynomials of the Heisenberg algebra generators \footnote{It is
indeed the universal enveloping of the Heisenberg algebra which in suitable
basis can be defined as the symmetrized products of the $\mathfrak{h}_d$
generators, owing the Poincar\'e-Birkhoff-Witt theorem (see eg. \cite{Fuchs}).}.

An associative algebra can always be endowed with a commutator product, thus yielding a Lie algebra,
\begin{equation}
[\mathfrak{h}_d^*]=\Big\{\mathfrak{h}_d^*\,; \quad [A,\,B]\,:=\,AB-BA,\quad
A,\;B\quad\in \quad \mathfrak{h}_d^* \Big\}\ .\label{WLA}
 \end{equation}
Alternatively, as the Weyl algebra is graded over $\mathbb{Z}_2$, we can endow their generators with a supercommutator product yielding a Lie superalgebra, 
\begin{equation}\label{WLSA}
 [\mathfrak{h}_d^*\}:=\Big\{\mathfrak{h}_d^*\,;\quad
[A,\,B\}\,:=\, A B-(-1)^{|A||B|} BA,\quad A,\;B\quad \in \quad \mathfrak{h}_d^* \Big\}.
\end{equation}
Here, $|\cdot|$ denotes the degree of the generators, respectively $|\cdot|=0$ and $|\cdot|=1$ for even and odd order polynomials in the generators $G_i$ and $P_i$ of the Heisenberg algebra. The constants of structure of \eqref{WLA} and \eqref{WLSA} are derived from those of $\mathfrak{h}_d$, and the (super)Jacobi identity follows from  the associativity of the Weyl algebra.     
 The algebras \eqref{WLA} and \eqref{WLSA} are called \textit{higher spin (super)algebras} \cite{Vasiliev1,Vasiliev2,Vasiliev3}, since they contain a maximal subalgebra of compact generators under which the remaining generators transform as tensors of arbitrary spin. 

In our approach the Heisenberg algebra is composed of a mass-central-charge, Galileo boosts and translations generators of a non-relativistic particle,
\begin{equation}
\mathfrak{h}_d=\{m=\mathds{1},\,G_i=x_i-tP_i,\; P_i=-{\rm i}\partial / \partial x_i\,
 \label{repGP1}\}.
\end{equation}
Consider now the free Schr\"odinger equation in $d$ spatial dimensions,
\begin{equation}\label{scheq}
\widehat{S} \,| \psi(x,t) \rangle =0, \qquad
\widehat{S}={\rm i}\frac{\partial}{\partial t}-H,
\qquad H=\frac12 \overrightarrow{P}\,{}^2.
\end{equation} 
The dynamics of a quantum operator is given by the Heisenberg equation $\dot{\mathcal{O}}=\partial {\cal O}/{\partial t}+ {\rm i} [H, {\cal O}],$ which can be written in terms of the Schr\"odinger operator as 
\begin{equation}\label{Heisenberg}
\dot{\mathcal{O}}=-{\rm i}\left[\widehat{S}, {\cal O}\right]\,.
\end{equation}
This equation can be also regarded as the first class constraint associated to the time-parametrization invariance of the free non-relativistic particle \cite{Gomis1}. 
 
The equations \eqref{scheq} and \eqref{Heisenberg} imply that constants of motion are symmetry generators, since, as they commutes with the Schr\"odinger operator, 
they leave invariant the space of solutions of the Schr\"odinger  equation. 

The product of constant of motion is also a constant of motion. This follows from the Leibniz rule satisfied by the derivative with respect to the time and by the adjoint action of $\widehat{S}$ acting on the product of constants of motion.
For the free Schrodinger equation, from Galileo boosts and translations
\begin{equation}
\dot{G}_i=-{\rm i}[\widehat{S}, \, G_i]=0,\qquad \dot{P}_i=-{\rm i}[\widehat{S},\, P_i]=0\,,\label{SGSP}
\end{equation}
any polynomial of $G_i$ and ${P}_i$ will be also a constant of motion. Therefore the free Schr\"odinger equation admits infinitely many conserved charges, spanned by arbitrary operator functions of $G_i$ and ${P}_i$, and which contains the Weyl algebra as the basis of the polynomial class of functions.

\section{From Galilean to higher spin symmetries}

Once the generators $G_i$ and $P_i$ are  provided we can form the vector
\begin{equation}\label{La}
L_a=(G_1,...,G_d,P_1,...,P_d), \qquad a=1,...,2d.
\end{equation} 
The commutation relations of $G_i=L_i$ and $P_j=L_{d+j}$ become now,
\begin{equation}\label{Cabn}
[L_a,L_b]={\rm i}C_{ab}, \qquad C_{ab}={\scriptsize
\left(%
\begin{array}{cc}
  0 & I_{d\times d} \\
  -I_{d\times d} & 0 \\
\end{array}
 \right)},
\end{equation}
which defines the symplectic matrix $C_{ab}$. The symmetrized second order products of generators \eqref{La}, 
\begin{equation}
M_{ab}=\frac{1}{2}\{L_a,L_b\},\label{Mab}
\end{equation}
commute with themselves as,
\begin{equation}
[ M_{ab},\, M_{cd} ]={\rm i}(C_{ac} M_{bd}+C_{bd}M_{ac}+C_{ad}M_{bc}+C_{bc}M_{ad}),\label{MabMcd}
\end{equation}
as it is deduced from \eqref{Cabn}, generating a representation of the
$\mathfrak{sp}(2d)$ algebra. This representation is usually referred as to
``oscillator representation'', since one of their compact generators can be
identified with a harmonic oscillator Hamiltonian. Here the Hamiltonian is, however, identified with a non-compact generator, the one of the
non-relativistic free particle. 

$M_{ab}$ together with $L_a$ yields the commutation relations, 
\begin{equation}
[M_{ab},L_c]={\rm i}(C_{ac}L_b+C_{bc}L_a).\label{MabLc} 
\end{equation}
 The generators $M_{ab}$, $L_a$ and $\mathds{1}$ yields the maximal finite dimensional subalgebra of \eqref{WLA} (see \eqref{Cabn}-\eqref{MabMcd}-\eqref{MabLc}), 
\begin{equation}\label{hdsp2d}
 \mathfrak{h}_d\niplus \mathfrak{sp}(2d) =\Big\{\mathds{1}, L_a,\,M_{ab}\,;\;[\cdot,\cdot] 
\Big\} \quad \subset \quad [\mathfrak{h}_d^*] \ .
\end{equation}
From $M_{ab}$ we can define the generators,
\begin{equation}
C=\frac{1}{2} \delta^{ij}M_{ij},
\quad D=-\delta^{ij}M_{i\,d+j}, 
\quad H=\frac{1}{2}\delta^{ij}M_{d+i\,d+j}\,, 
\quad
J_{ij}= M_{i\,d+j}-M_{j\,d+i}, \quad i,j=1,...,d\ .\label{calCDHJ} 
\end{equation}
Now \eqref{calCDHJ} together with \eqref{repGP1} yields the Schr\"odinger algebra $\mathfrak{sch}(d)$, 
\begin{eqnarray}
 &&[G_i,\, P_j] =  {\rm i} m \delta_{ij}\,,\label{GP} \nonumber\\ 
 &&[J_{ij},\, J_{kl}] = {\rm i}( \delta_{ik} J_{jl} + 
   \delta_{jl} M_{ik} - \delta_{il} J_{jk} - \delta_{jk} J_{il}), \label{rot1} \nonumber\\
  &&[J_{ij},\, P_k] = {\rm i}(\delta_{ik} P_j - \delta_{jk} P_i), \quad
  [J_{ij},\, G_k] = {\rm i}(\delta_{ik} G_j - \delta_{jk} G_i), \label{vect}\nonumber\\
  &&[H,\, G_i] = -{\rm i} P_i,\nonumber\\
 &&[D,\,C]=2{\rm i}C,\qquad [D,\,H]=-2{\rm i}H,\qquad [H,\,C]={\rm i}D, \label{CDH}\nonumber\\ 
&&[D,\, P_i] = -{\rm i} P_i, \quad [D,\, G_i] = {\rm i} G_i,\qquad [C,\, P_i] =  {\rm i} G_i\,.\label{DPDG}\nonumber
\end{eqnarray}
Other commutators vanish. We stress that the full Schr\"odinger algebra $\mathfrak{sch}(d)$ is implied by the Heisenberg commutation relation \eqref{Cabn} and the definitions \eqref{calCDHJ}.
Indeed, from \eqref{repGP1} and \eqref{calCDHJ}, the standard representation of  $\mathfrak{sch}(d)$ is recovered,
\begin{eqnarray}
&&\hbox{Hamiltonian }\quad H=\overrightarrow{P}\,{}^2 /2,\label{H}\nonumber\\
&&\hbox{rotations }\quad J_{ij}=x_i P_j-x_j P_i,\label{rot}\nonumber\\
&&\hbox{translations }\quad P_i=-{\rm i}\partial / \partial x_i,\label{trans}\nonumber\\
&&\hbox{boosts }\quad G_i=mx_i-tP_i,\label{boost}\nonumber\\
&&\hbox{mass } \quad m=\mathds{1},\label{mass}\nonumber\\
&&\hbox{expansions } \quad C=-t^2 H+tD+\overrightarrow{x}\,{}^2/2,\qquad \label{Crep}\nonumber\\
&&\hbox{dilatations }\quad D=2tH-\overrightarrow{x}\cdot \overrightarrow{P}+{\rm i}d/2 \, .\label{Drep}\nonumber
\end{eqnarray}
The Schr\"odinger algebra, having the structure $\mathfrak{sch}(d) = \mathfrak{h}_d \niplus \{\mathfrak{so}(d) \oplus \mathfrak{sl}(2,\mathbb{R})\},$ is subalgebra of  \eqref{hdsp2d} and the higher spin algebra \eqref{WLA}, i.e.  $\mathfrak{sch}(d)\subset (\mathfrak{h}_d\niplus \mathfrak{sp}(2d) ) \subset [\mathfrak{h}_d^*]$. 

Rotations, dilations and expansion generators, are second order operators in Galileo boosts and translations.  They generate independent symmetries however, as it is well known. Indeed, the finite transformations generated by $D$ and $C$ are respectively (see e.g. \cite{Nie,Duval:2009vt}) $(t,x_i)\; \rightarrow \; (\lambda^2  t, \lambda x_i)$ and $(t,x_i)\; \rightarrow \; (1-\kappa t)^{-1} (t, x_i)$ where $\lambda$ and $\kappa$ are transformation parameters.

The new generators of $\mathfrak{h}_d\niplus \mathfrak{sp}(2d)$ not contained in the Schr\"odinger algebra are of second order in spatial derivatives,
\begin{equation}
\begin{array}{l}
P_{ij}=\frac12 M_{d+i\,d+j}-\frac1d \delta_{ij} H,\quad
G_{ij}=\frac12 M_{ij}-\frac1d \delta_{ij} C,\quad
Z_{ij}=M_{i\,d+j}+M_{j\,d+i}+\frac2d \delta_{ij} D, \\[6pt]
P_{ij}= P_iP_j-\frac1d \delta_{ij} H,\quad
G_{ij}=G_iG_j-\frac1d \delta_{ij} C,\quad
Z_{ij}=G_iP_j+P_iG_j+\frac2d \delta_{ij} D. 
\end{array}
\label{PGZ}
\end{equation}
These operators are traceless, $\delta^{ij}P_{ij}=\delta^{ij}G_{ij}=\delta^{ij}Z_{ij}=0.$
 The non-vanishing remaining commutation relations read,
\begin{equation}
 \begin{array}{l}
 [P_{ij},\,G_k]=-\frac{{\rm i}}{2}(\delta_{ik}P_j+\delta_{jk}P_i)+\frac{{\rm i}}{d}\delta_{ij}P_k,\quad [Z_{ij},\,G_k]=-{\rm i}(\delta_{ik}G_j+\delta_{jk}G_i)+\frac{2{\rm i}}{d}\delta_{ij}G_k\,,
 \\[6pt]
 [G_{ij},\,P_k]=\frac{{\rm i}}{2}(\delta_{ik}G_j+\delta_{jk}G_i)+\frac{{\rm i}}{d}\delta_{ij}G_k,\quad [Z_{ij},\,P_k]={\rm i}(\delta_{ik}P_j+\delta_{jk}P_i)-\frac{2{\rm i}}{d}\delta_{ij}P_k \,,\\[6pt]
 [H,\,G_{ij}]=-\frac{{\rm i}}{2}Z_{ij},\qquad [H,\,Z_{ij}]=-4{\rm i} P_{ij},\qquad [C,\,P_{ij}]=\frac{{\rm i}}{2}Z_{ij},\qquad [C,\,Z_{ij}]=4{\rm i} G_{ij}\,,
\\[6pt]
 [D,\,G_{ij}]=2{\rm i}G_{ij},\qquad [D,\,P_{ij}]=-2{\rm i} P_{ij},\\[6pt]
[G_{ij},\,P_{kl}]=\frac{{\rm i}}{8}(\delta_{ik}(Z_{jl}+J_{jl})+\delta_{il}(Z_{jk}+J_{jk})+\delta_{jk}(Z_{il}+J_{il})+\delta_{jl}(Z_{ik}+J_{ik}))\\[6pt]
 \qquad \qquad \quad-\frac{{\rm i}}{2d}(\delta_{ik}\delta_{jl}+\delta_{il}\delta_{jk}-\frac{2}{d}\delta_{ij}\delta_{kl})D- \frac{{\rm i}}{2d}(\delta_{ij} Z_{kl}+\delta_{kl} Z_{ij})\,,\\[6pt]   
 [G_{ij},\,Z_{kl}]= {\rm i}(\delta_{ik}G_{jl}+\delta_{il}G_{jk}+\delta_{jk}G_{il}+\delta_{jl}G_{ik})\\[6pt]
 \qquad \qquad \quad+\frac{2{\rm i}}{d}(\delta_{ik}\delta_{jl}+\delta_{il}\delta_{jk}-\frac{2}{d}\delta_{ij}\delta_{kl})C 
-\frac{4{\rm i}}{d}(\delta_{ij} G_{kl}+\delta_{kl} G_{ij}),\\[6pt]   
 [P_{ij},\,Z_{kl}]= -{\rm i}(\delta_{ik}P_{jl}+\delta_{il}P_{jk}+\delta_{jk}P_{il}+\delta_{jl}G_{ik})\\[6pt]
 \qquad \qquad \quad-\frac{2{\rm i}}{d}(\delta_{ik}\delta_{jl}+\delta_{il}\delta_{jk}-\frac{2}{d}\delta_{ij}\delta_{kl})H+ \frac{4{\rm i}}{d}(\delta_{ij} P_{kl}+\delta_{kl} P_{ij}) ,\\[6pt] 
 [Z_{ij},\,Z_{kl}]= -2{\rm i}(\delta_{ik}J_{jl}+\delta_{il}J_{jk}+\delta_{jk}J_{il}+\delta_{jl}J_{ik}),\label{ZijZkl}
\end{array}\nonumber
\end{equation}
extending $\mathfrak{sch}(d)$ to  $\mathfrak{h}_d\niplus \mathfrak{sp}(2d$),
which in Galileo covariant notation is generated by
\footnote{The algebra \eqref{hd22nr} has been also discussed in a different context \cite{Castro:2009en}.} (cf. \eqref{hdsp2d}),
\begin{equation}\label{hd22nr}
\mathfrak{h}_d\niplus \mathfrak{sp}(2d) =\Big\{\mathds{1},\,G_i,\, P_i,\,J_{ij},\, C,\,D,\,H,\,P_{ij},\,  G_{ij},\, Z_{ij}\,;\;[\cdot,\cdot]
\Big\}.
\end{equation}
Notice that endowing the generators \eqref{hd22nr}  instead with a
supercommutator product, which is possible owing the $\mathbb{Z}_2$ grading
\eqref{WLSA} \footnote{Indeed, it is the reflection operator,
$R\Psi(x)=\Psi(-x)$, which induces the $\mathbb{Z}_2$ grading, as it 
anti-commutes with
Galileo boosts, translations and all their odd powers in the Weyl algebra,
whereas it commutes with their even order
powers. The symmetric and
antisymmetric projections of the wave function $\Psi_\pm(x,t)=\frac12
(\Psi(x,t)\pm \Psi(-x,t))$, can be seen as they were ``bosonic''  and
``fermionic''. Observe also that the odd projection satisfies a Pauli
exclusion like principle, $\Psi_-(0,t)=0$. }, yields the
superalgebra 
\begin{equation}\label{osp12du1} 
\mathfrak{osp}(1|2d)\oplus \mathfrak{u}(1)=\Big\{\mathds{1},\,G_i,\, P_i,\,J_{ij},\, C,\,D,\,H,\,P_{ij},\,  G_{ij},\, Z_{ij}\,;\;[\cdot,\cdot\} \Big\}\ ,
\end{equation}
with (anti)commutation relations equivalent to \eqref{Mab}, \eqref{MabMcd} and \eqref{MabLc} (see the definitions \eqref{La}, \eqref{calCDHJ} and \eqref{PGZ}). It is the maximal finite dimensional subalgebra of \eqref{WLSA}. Here, the Galileo boosts and translations generators, $G_i$ and $P_i$, are regarded as supercharges (cf. \eqref{PGZ}), 
\begin{eqnarray}
\{G_i,\, G_j\} = 2G_{ij}+\frac2d\delta_{ij}C,\quad \{P_i,\, P_j\} = 2P_{ij}+\frac2d\delta_{ij}H,\quad \{G_i,\, P_j\} = J_{ij}+Z_{ij}-\frac2d\delta_{ij}D . \label{antiGP}\nonumber
\end{eqnarray}
Indeed, the vanishing commutation relations of the Hamiltonian with the non-trivial reflection grading-operator reveals the double degeneracy of the Hamiltonian own to supersymmetry, which is characteristic in supersymmetric quantum mechanics.

The Schr\"odinger algebra extension \eqref{hdsp2d} can be seen hence as the
bosonic counterpart of the orthosymplectic supersymmetry $\mathfrak{osp}(1|2d)$
, using the terminology of \cite{Fed} where the bosonic counterpart of
super-Poincar\'e was studied.
Of course, the introduction of other degrees of freedom such as Clifford or
Grassmann variables as in, e.g., \cite{Duval:1993hs,Gauntlett:1990xq}, would
yield a more standard type of supersymmetry.
The supersymmetry  induced by parity under spatial reflections has
been widely studied by M. Plyushchay and collaborators in diverse interacting
quantum mechanical system which do not involve fermionic degrees of freedom, and
called for this reason \textit{bosonized supersymmetry}, or \textit{hidden
supersymmetry} \cite{bsusy}.

\subsection*{Non-relativistic covariance of Vasiliev equations}
 
The theory of Vasiliev is a generalization of the Cartan formulation of
gravity (see e.g. \cite{Vasiliev3}), determining the dynamics of the higher spin
fields forms by means of a Cartan-integrable system of equations. Topological higher spin gravity admits also a Chern-Simons action principle \cite{Blencowe} (see also \cite{Henneaux} in a more recent context) in three dimensions, and in four dimensions an action principle was proposed in
\cite{Boulanger}. 
Here we show that the higher spin gauge theory \cite{Vasiliev3,Vasiliev4,Blencowe} exhibits also non-relativistic symmetries.

 Vasiliev's theory makes use of differential forms valued in the higher spin
gauge algebra of the type
\begin{equation}\label{W}
W(X)=\sum_{n=0}^{\infty} \frac{1}{n!}\,  W{}^{a_1 a_2 \cdots
a_n}(X) \ Y_{a_1}Y_{a_2}\cdots Y_{a_n},\quad a_1, a_2,...,a_n=1,2,...,2d.
\end{equation}
 $W^{a_1 a_2 \cdots a_n}(X)$ are completely
symmetric in $a$-indices and $X$ labels local coordinates of the base space-time manifold. 
The gauge field \eqref{W} may involve also the reflection operator (or Klein
operator) in its expansion, as well as Clifford or Grassmann variables, which we
omit here for simplicity. The representation of the
Weyl algebra can be realized also in terms of commuting $Y_a$-symbols endowed with an
associative star product, whereas the higher spin (super)algebra is obtained
from the correspondent (anti)symmetrization of the star product. Identifying the 
Heisenberg algebra of the oscillators $Y_a$ with the 
Galileo-boost/translation generators \eqref{La} (see Table \ref{T1}),
 allow us to view the expansion \eqref{W} as a one-form
valued in the universal enveloping of the non-relativistic
Schr\"odinger algebra $\mathfrak{h}_d^*$, making explicit its non-relativistic
covariance. 
\begin{table}[ht]
\begin{center}
\begin{tabular}{|c|c|c|c|}
 \hline
\begin{tabular}{c}
 HS theory \\
dimension
\end{tabular} & \begin{tabular}{c}
 notation for Heisenberg \\
algebra generators 
\end{tabular} & \begin{tabular}{c}
 dimension\\
NR theory 
\end{tabular}
&\begin{tabular}{c}
 Correspondence with\\
NR generators 
\end{tabular} \\\hline\hline
  $2+1$ & $Y_a, \, a=1,2$ & $1+1$ &
$Y_1\rightarrow G_1,\,Y_2\rightarrow P_1$  \\[5pt]\hline
  $3+1$ & $Y_a=(y_\alpha,y_{\dot{\alpha}}) \, \alpha,\dot{\alpha}=1,2$ & $2+1$ &
$Y_\alpha\rightarrow(G_1,P_1)\,,\,
Y_{\dot{\alpha}}\rightarrow(G_2,P_2)$\\[7pt]\hline
$D>4$ & 
\begin{tabular}{c}
$Y^A_a\,,$\\[4pt]
$a=1,2\,,\,A=0,...,D-1,$
\end{tabular} & $D+1$ &
\begin{tabular}{c}
$Y^0_a\rightarrow(P_1,G_1),\, Y^i_a\rightarrow(G_{i+1},P_{i+1}),$ \\[4pt]
 $i=1,...,D-1$
\end{tabular}
\\\hline
\end{tabular}
\end{center}
\caption{The non-relativistic covariance of Vasiliev's higher spin theory can
be made explicit identifying the Heisenberg algebra generators $Y$'s with
the generators of Galilean boost and translations. 
}
\label{T1}
\end{table}

In $D>4$ \cite{Vasiliev4} the Heisenberg
algebra employed in Vasiliev's theory is given by $[Y^A_a,Y^B_b]={\rm i}\, C_{ab}\,\eta^{AB}$, where
$\eta^{AB}$ is the Lorentz metric in
$D$ dimensions. Thus the formulation of higher spin gravity in general (relativistic) space-time dimensions $D>4$ makes use of Heisenberg algebras generated by vectors $Y^A_a$ ($A=0,...,D-1$, $a=1,2$),  so the internal phase-spaces have dimension $2\times D$. In three and four dimensions these phase-spaces have respectively dimensions $2$ and $4$, and the generators of the Heisenberg algebras transform as spinors under the relativistic Lorentz group. Therefore the latter phase-spaces can be labeled instead in terms of Galilean non-relativistic covariance in dimension $D+1$, including the time direction, for higher-spin gravity in $D>4$, while the non-relativistic covariance is $1+1$ (one spatial non-relativistic direction and the time) for higher-spin gravity in $D=3$ dimensions, and it is $2+1$ (two non-relativistic spatial direction and the time) for higher-spin gravity in $D=4$ relativistic space-time. The expected correspondence between higher spin gravity and non-relativistic quantum mechanics is therefore as given in the table \ref{T1}.

 \section{Conclusions}

We have shown in a simple way how the free Schr\"odinger equation enjoys infinite symmetries generated by the Weyl algebra. Since the Weyl algebra can be made covariant under non-relativistic (Galilean) transformations and also under relativistic (higher spin) transformations, we observe that both, the Schr\"odinger equation and higher spin gravity, have non-relativistic and relativistic symmetries, depending on the choice of algebra labels. 

More generally, since the (unitary) representations
of the $\mathfrak{sp}(2d)$ algebra, or the $\mathfrak{osp}(1|2d)$ superalgebra,
admit the embeddings 
\begin{equation}\label{cfsp}\nonumber
\mathfrak{cf}(D-2,1)\approx \mathfrak{so}(D-1,2)
\qquad  \hookrightarrow \qquad \mathfrak{osp}(1|2^{[D/2]})\,. 
\end{equation}
the space of solution of the the free Schr\"odinger equation in $d=2^{[D/2]-1}$
spatial dimensions spans a representation of the (super)conformal algebra
in $D-1$ space-time dimensions, or the anti-de-Sitter algebra in $D$ dimensions.    

From our results, we would expect that Vasiliev theory in its complete formulation could apply also to non-relativistic
system.  It would be challenging to find these systems since they will be so closely related to gravity. Therefore, extensions of our study could lead to an holographic correspondence between higher spin theory and a non-relativistic quantum theory.
 
It is worth to mention here that the
spin-statistics theorem does not hold in non-relativistic field theories
\cite{Levy}, i.e. statistics and spin may be unconnected. Hence, there is not a
priory a reason to discard super-commutator product of the non-relativistic
particle symmetry generators in a first- or a second-quantized theory (cf.
\cite{Ramond}). In that case, one may speculate about the existence of a
new type of holographic correspondence, between supersymmetric relativistic
theories and non-relativistic theories which apparently are
non supersymmetric but which exhibit a $\mathbb{Z}_2$ graded structure.

The results here presented can be also generalized to the harmonic oscillator Schr\"odinger equation, taking advantage of the isomorphism of the Schr\"odinger algebra and the conformal Newton-Hooke algebra \cite{Nie}. We would like to thank the referee for pointing out that, more generally, there is a local diffeomorphism between solutions of linear second-order differential equations which in particular may be useful to map solutions of the free particle Schrodinger equation to the solutions of any second order Hamiltonian and \textit{vice versa}. See for instance \cite{Arnold}. Bearing in mind this theorem, we would expect that this theorem might be helpful to extend our results to any second order hamiltonian system in higher dimensions, at least locally.

\section*{Acknowledgments}

We thank X. Bekaert, M. Hassaine, P. Horvathy, M. Plyushchay and M.A. Vasiliev
for valuable discussions. This work was supported by  \textit{Anillos de
Investigaci\'on en Ciencia y Tecnolog\'ia}, project ACT 56 \textit{Lattice and
Symmetry} (Chile) and a CNRS (France) postdoctoral grant (contract number
87366).

\end{document}